%% file: main.tex
\documentclass[conference,comsoc]{IEEEtran}
\IEEEoverridecommandlockouts

\usepackage{./share/package_macros}
\input{./share/acronyms}

\title{Feedback-Aware Precoding for Millimeter Wave Massive MIMO Systems}

 \author{\IEEEauthorblockN{
 		Reza Ghanaatian\IEEEauthorrefmark{1},
 		Vahid Jamali\IEEEauthorrefmark{2},
 		Andreas Burg\IEEEauthorrefmark{1},
 		and Robert Schober\IEEEauthorrefmark{2}
 		}
 		\\
     \begin{tabular}[t]{c@{\extracolsep{1.5cm}}c} 
	      	\IEEEauthorrefmark{1}EPFL, Lausanne, Switzerland  & \IEEEauthorrefmark{2}FAU, Erlangen, Germany\\
	      	\small
	      	Email: \{\href{mailto:reza.ghanaatian@epfl.ch}{reza.ghanaatian}, 
	      	\href{mailto:andreas.burg@epfl.ch}{andreas.burg}\}@epfl.ch
	      	&
	      	\small
	      	Email: \{\href{mailto:vahid.jamali@fau.de}{vahid.jamali},
	      	\href{mailto:robert.schober@fau.de}{robert.schober}\}@fau.de
	      	\\ 
     \end{tabular}
 		      
} 

\begin{document} 
\maketitle	

\input{./sections/Abstract}
\input{./sections/Intro}

\input{./sections/SysMod}

\input{./sections/Feedback2}

\input{./sections/SimulationResults}
\input{./sections/Conclution}


\bibliographystyle{IEEEtran}
\bibliography{./share/IEEEabrv,./share/mmWaveJabref,./share/ConfAbrv}

\end{document}

%% file: share/acronyms.tex
\newacronym{MIMO}{MIMO}{multiple-input multiple-output}
\newacronym{CSI}{CSI}{channel state information}
\newacronym{CSIT}{CSIT}{channel state information at the transmitter}
\newacronym{SVD}{SVD}{singular value decomposition}
\newacronym{mmWave}{mmWave}{millimeter wave}
\newacronym{OMP}{OMP}{Orthogonal Matching Pursuit}
\newacronym{PS}{PS}{phase shifter}
\newacronym{AltMin}{AltMin}{alternating minimization}
\newacronym{AoD}{AoD}{angle of departure} 
\newacronym{AoA}{AoA}{angle of arrival}
\newacronym{SNR}{SNR}{signal-to-noise ratio}
\newacronym{BER}{BER}{bit error rate}
\newacronym{MMSE}{MMSE}{minimum mean square error}
\newacronym{QPSK}{QPSK}{quadrature phase shift keying}
\newacronym{FDD}{FDD}{frequency division duplex}
\newacronym{LTE}{LTE}{Long-Term Evolution}

%% file: sections/Abstract.tex
\begin{abstract}
\Gls{mmWave} communication is a promising solution for coping with the ever-increasing mobile data traffic because of its large bandwidth.
To enable a sufficient link margin, a large antenna array employing directional beamforming, which is enabled by the availability of \gls{CSIT}, is required. 
However, \gls{CSIT} acquisition for \gls{mmWave} channels introduces a huge feedback overhead due to the typically large number of transmit and receive antennas. 
Leveraging properties of \gls{mmWave} channels, this paper proposes a precoding strategy which enables a flexible adjustment of the feedback overhead.
In particular, the optimal unconstrained precoder is approximated by selecting a variable number of elements from a basis that is constructed as a function of the transmitter array response, where the number of selected basis elements can be chosen according to the feedback constraint.
Simulation results show that the proposed precoding scheme can provide a near-optimal solution if a higher feedback overhead can be afforded.
For a low overhead, it can still provide a good approximation of the optimal precoder.
	
\end{abstract}


%% file: sections/Intro.tex
\section{Introduction}\label{sec:intro}

Communication over millimeter wave (mmWave) frequencies is defining a new era of wireless communication, as it can enable gigabit-per-second data rates because of the large available channel bandwidth. In order to reap the benefits of \gls{mmWave} communication, however, massive antenna arrays need to be employed, which provide a sufficient beamforming gain to combat the high path loss at \gls{mmWave} frequencies.

In traditional \gls{MIMO} systems, fully digital processing is enabled by connecting each antenna to the baseband processor. In a practical \gls{mmWave} system employing a large number of antennas, i.e., a massive \gls{MIMO} system, this approach is not feasible anymore due to the high cost and high power consumption of \gls{mmWave} mixed signal and RF components. Hybrid \gls{MIMO} architectures \cite{han2015large, zhang2005variable} overcome this limitation by utilizing only few RF chains and by processing the signals in both the digital and analog domains.
The digital baseband processing provides full control over both the phase and the amplitude of the signal, while the analog RF processing enabled by \glspl{PS} can only control the phase of the signal, i.e., there is a constant modulus constraint.
Despite the lower cost and power consumption of hybrid architectures, they impose a set of constraints on \gls{MIMO} precoding/combining due to the limitations imposed by analog processing. Several approaches have been proposed for precoding/combining in such architectures \cite{el2014spatially,yu2016alternating, rusu2016low}. In \cite{el2014spatially}, the sparse structure of the \gls{mmWave} channel is exploited to formulate the precoding problem and then the \gls{OMP} algorithm is used to approximate the optimal unconstrained (digital) precoder, which maximizes the achievable rate. In \cite{yu2016alternating}, hybrid precoding is treated as a matrix factorization problem and \gls{AltMin} algorithm is used to find a solution. The authors of \cite{rusu2016low} propose several low-complexity solutions that provide different tradeoffs between precoding performance and algorithm complexity.

Any precoding algorithm relies on the availability of \gls{CSI} at the transmitter.
For \gls{FDD} systems, which are a popular choice, e.g., for \gls{LTE}, and the focus of this paper, channel reciprocity does not hold and hence the \gls{CSI} can be acquired at the transmitter only via an explicit feedback link.
The existing \gls{CSI} feedback strategies for MIMO systems can be classified into two categories: i) \emph{codebook-based} feedback, where an index from a known codebook is fed back, and ii) \emph{individual quantized} feedback, where each \gls{CSI} value is quantized and send back~\cite{love2004value}. 
Both approaches have shortcomings when it comes to systems with large antenna arrays \cite{wang2017hybrid}, such as mmWave systems, and thus the limitation of the feedback channel should be taken into account for the design of precoding algorithms \cite{hur2013millimeter,noh2017multi,alkhateeb2014channel,alkhateeb2015limited,Ren2017}. In \cite{noh2017multi} and \cite{alkhateeb2014channel}, multi-resolution codebooks are proposed to minimize the training power and feedback overhead during channel estimation. The authors of \cite{alkhateeb2015limited} and \cite{Ren2017} propose a limited feedback precoding scheme for a multi-user scenario, where the RF precoder is configured by using codebook-based feedback from each user. The baseband precoder is then computed based on the effective channels of the users, which have a much lower dimension than the complete channel matrix, and thus can be quantized and fed back to the transmitter. 
While the above approaches are trying to optimize and reduce the feedback overhead required for precoder computation at the transmitter, the tradeoff between precoder performance and feedback rate is unclear and not easily adjustable.

In this paper, we propose a feedback-aware strategy for hybrid precoder design by exploiting the properties of the \gls{mmWave} channel and the transmitter architecture introduced in~\cite{zhang2014achieving} that employs two \glspl{PS} per RF chain for each antenna and is capable of realizing any arbitrary precoder via hybrid precoding.\footnote{We note that the hybrid MIMO architecture in~\cite{zhang2014achieving} employs twice the number of \glspl{PS} as the widely-adopted fully-connected hybrid architecture in~\cite{el2014spatially} and \cite{rusu2016low}. However, the minimum number of RF chains (i.e., the number of data streams) is used in~\cite{zhang2014achieving}, whereas the number of RF chains in a fully-connected architecture can be in general equal to or larger than the number of data streams. Nevertheless, \glspl{PS} at \gls{mmWave} frequencies can be manufactured using, e.g., simple RF delay lines~\cite{balanis1992antenna}. In this paper, we exploit the additional degrees of freedom that the MIMO architecture in~\cite{zhang2014achieving} provides to design a feedback-aware precoder.}
In particular, we propose to approximate the optimal unconstrained precoder by selecting a variable number of vectors from a basis matrix that is constructed as a function of the transmitter array response, where the number of selected basis vectors can be adapted according to the feedback constraint.
We further propose a multi-beam basis matrix that improves the approximation accuracy especially when the affordable feedback overhead is low.
Finally, we characterize the total feedback overhead for our proposed scheme and we provide a comparison with state-of-the-art precoder designs.
We show that the proposed precoder design offers a flexible tradeoff between performance and feedback overhead.
In particular, the optimal precoder can be approximated with high accuracy at the expense of an increased feedback overhead, whereas by limiting the feedback overhead, an acceptable approximation of the optimal precoder can still be maintained.

\emph{Notation:} Throughout this paper, $\mathbb{C}$ is the field of complex numbers; 
$\mathbb{A}$ is the field of modulo-one complex numbers;
bold upper-case letters $\mat A$ denote matrices;
bold lower-case letters $\mat a$ denote vectors; 
lower-case letters $a$ are scalars;
$\mat A_{i,j}$ is the entry in the $i$-th row and $j$-th column of $\mat A$;  
$\mat A\hermis$ and $\mat A\transp$ denote conjugate transpose and transpose of matrix $\mathbf{A}$, respectively; 
$\frob {\mat A}$, $|\mat A|$, $|a|$, and $\angle a$ are the Frobenius norm of $\mat A$, the determinant of $\mat A$, the magnitude of $a$, and the phase of $a$, respectively; 
$\mathbf{I}_N$ is the $N \times N$ identity matrix;  $\mathbf{0}_{N \times M}$ is the $N \times M$ all-zero matrix;
$\mathcal{CN}(m,\sigma^2)$ denotes the complex normal distribution with mean $m$ and variance $\sigma^2$;
and expectation is denoted by $\Ex \cdot$.

%% file: sections/SysMod.tex
\section{System Model and Problem Statement}\label{sec:sysmod}

In this section, we first introduce the considered system model. Subsequently, we formally present the problem statement for the feedback-aware precoder design.

\subsection{System Model}

We consider a narrow-band point-to-point MIMO system consisting of an $M$-antenna transmitter that sends $S$ independent data streams to an $N$-antenna receiver. The input-output channel model is given by
\begin{IEEEeqnarray}{lll} \label{Eq:AWGN}
\mathbf{y} = \mathbf{H}\mathbf{F}\mathbf{s}+\mathbf{z},
\end{IEEEeqnarray}
where $\mathbf{y}\in\mathbb{C}^{N\times 1}$ denotes the received signal vector and $\mathbf{s}\in\mathbb{C}^{S\times 1}$ represents the vector of transmitted independent data streams. We assume $\Ex{\mathbf{ss}\hermis}=P\mathbf{I}_{S}$, where $P$ denotes the transmit power. Moreover, $\mathbf{z}\in\mathbb{C}^{N\times 1}$ denotes the additive white Gaussian noise vector at the receiver, i.e., $\mathbf{z}\sim\mathcal{CN}(\mathbf{0}_{N \times 1},\sigma_n^2\mathbf{I}_N)$ where $\sigma_n^2$ is the noise variance at each receive antenna. In (\ref{Eq:AWGN}),  $\mathbf{F}\in\mathbb{C}^{M\times S}$ represents the precoding matrix satisfying $\|\mathbf{F}\|_F=1$ and $\mathbf{H}\in\mathbb{C}^{N\times M}$ denotes the channel matrix, which are both discussed in more detail in the following.

\subsubsection{Sparse \gls{mmWave} Channel Model}
\gls{mmWave} channels are expected to have limited scattering due to the high free-space path loss. To incorporate this effect, a clustered channel model, i.e., the Saleh-Valenzuela model, is considered \cite{rappaport2014millimeter}. In this model, the channel is assumed to be a superposition of $L$ scattering clusters, each of which contributes $J$ rays (i.e., in total they are $LJ$ paths) as
\begin{IEEEeqnarray}{lll} \label{Eq:Channel}
	\mathbf{H} = \sqrt{\frac{MN}{LJ}}\sum_{k=1}^{LJ} h_{k} \mathbf{h}_r(\theta_{k})\mathbf{h}_t\hermis(\phi_{k}),
\end{IEEEeqnarray}
where $h_{k}\in\mathbb{C}$ is the channel coefficient of the $k$-th path and $\mathbf{h}_t(\phi_{k})$ ($\mathbf{h}_r(\theta_{k})$) denotes the transmitter (receiver) antenna array response vector at \gls{AoD} (\gls{AoA}) $\phi_{k}$ ($\theta_{k}$), assuming a horizontal beamforming array only.
For a uniform linear array, we obtain $\mathbf{h}_t(\phi)$ and $\mathbf{h}_r(\theta)$ as
\begin{IEEEeqnarray}{lll} \label{Eq:ULA}
	\mathbf{h}_t(\phi) &= \frac{1}{\sqrt{M}}\left[1,e^{j\frac{2\pi d}{\lambda}\sin(\phi)},\dots,e^{j(M-1)\frac{2\pi d}{\lambda}\sin(\phi)}\right]\transp, \; \IEEEyesnumber\IEEEyessubnumber \\
	\mathbf{h}_r(\theta) &= \frac{1}{\sqrt{N}}\left[1,e^{j\frac{2\pi d}{\lambda}\sin(\theta)},\dots,e^{j(N-1)\frac{2\pi d}{\lambda}\sin(\theta)}\right]\transp, \IEEEyessubnumber
\end{IEEEeqnarray}
where $\lambda$ denotes the wave length and $d$ is the spacing between the antennas. 

\begin{figure}[t]
	\centering
	\includegraphics[width=\linewidth]{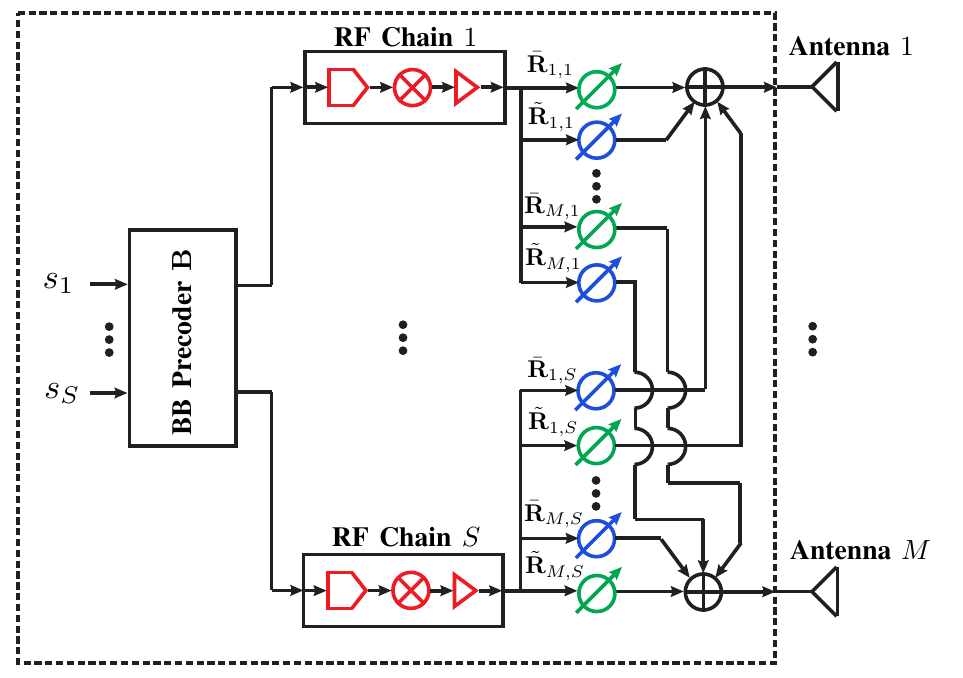}
	\caption{Schematic illustration of the considered massive MIMO architecture consisting of $M$ antennas, $S$ RF chains (i.e., digital-to-analog converters, mixers, and signal amplifiers (shown in red)), $2MS$ \glspl{PS} (i.e., two \glspl{PS} per RF chain per antenna (shown in green and blue, respectively)), and a baseband (BB) precoder \cite{rappaport2014millimeter}.}
	\vspace{-3mm}
	\label{fig:SysMod}
\end{figure}

\subsubsection{Hybrid Precoder}
In this paper, we consider the hybrid MIMO architecture introduced in \cite{zhang2014achieving} and \cite{bogale2016number}, which is illustrated in Fig.~\ref{fig:SysMod}. This hybrid architecture consists of $S$ RF chains and $2M$ \glspl{PS} per RF chain and has the advantage of being able to represent any fully digital precoder. The following lemma provides the structure of $\mathbf{F}$ for this hybrid MIMO architecture.

\begin{lem}
Any arbitrary precoder $\mathbf{F}\in\mathbb{C}^{M\times S}$ can be decomposed as
\begin{IEEEeqnarray}{lll} \label{Eq:Hybrid}
\mathbf{F} = \mathbf{R}\mathbf{T}\mathbf{B},
\end{IEEEeqnarray}
where $\mathbf{T}=[\mathbf{I}_S,\mathbf{I}_S]\transp$. Here, $\mathbf{B}\in\mathbb{C}^{S\times S}$ is the baseband precoder with entries 
\begin{IEEEeqnarray}{lll} 
\mathbf{B}_{s,s'} = \begin{cases}
\frac{1}{2} \underset{1\leq m \leq M}{\max} \,\,|\mat F_{m,s}|,\quad &\mathrm{if}\,\, s=s'\\
0, &\mathrm{otherwise,}
\end{cases}
\end{IEEEeqnarray}
where $s,s'=1,\dots, S$. Diagonal matrix $\mathbf{B}$ can be implemented by $S$ RF chains. Furthermore, $\mathbf{R}\in\mathbb{A}^{M\times 2S}$ is the RF precoder which can be written as $\mathbf{R}=[\bar{\mathbf{R}},\tilde{\mathbf{R}}]$, where the entries of $\bar{\mathbf{R}}\in\mathbb{A}^{M\times S}$ and $\tilde{\mathbf{R}}\in\mathbb{A}^{M\times S}$ are given by
\begin{IEEEeqnarray}{lll} 
\bar{\mathbf{R}}_{m,s} =\mathrm{exp}\left(j\left[\angle(\mathbf{F}_{m,s})+\cos^{-1}\left(\frac{|\mathbf{F}_{m,s}|}{2\mathbf{B}_{s,s}}\right)\right]\right), 
\IEEEyesnumber\IEEEyessubnumber \\
\tilde{\mathbf{R}}_{m,s} =\mathrm{exp}\left(j\left[\angle(\mathbf{F}_{m,s})-\cos^{-1}\left(\frac{|\mathbf{F}_{m,s}|}{2\mathbf{B}_{s,s}}\right)\right]\right).\quad
\IEEEyessubnumber
\end{IEEEeqnarray}
Matrix $\mathbf{R}$ has modulo-one entries and hence can be implemented by a network of $2MS$ phase shifters.
\end{lem}
\begin{IEEEproof}
The proof is given in \cite[Theorem~1]{zhang2014achieving}. We note that the definitions of matrix $\mathbf{R}$ and $\mathbf{T}$ in~\cite{zhang2014achieving} are slightly different from those in this paper, cf. \cite[Eqs. (11)-(15)]{zhang2014achieving}. In fact, for notational simplicity, we decompose matrix $\mathbf{R}$ into $\bar{\mathbf{R}}$ and $\tilde{\mathbf{R}}$. As a result, matrix $\mathbf{T}$ consist of two identity matrices.
\end{IEEEproof}

\subsection{Problem Statement}

The hybrid architecture shown in Fig.~\ref{fig:SysMod} can achieve the same transmission rate as any fully-digital precoder, while the complexity is highly reduced because only $S$ RF chains are used. Although \emph{any} precoder matrix can be realized by the above decomposition strategy, the main problem for the design of \emph{general} precoders for massive MIMO systems is the feedback overhead from the receiver to the transmitter, which is either the full channel matrix $\mat H$, i.e., $M N$ complex-valued numbers, or the designed precoder $\mat F$, i.e., $M S$ complex-valued numbers. As this overhead scales with the number of transmit antennas, it becomes the \gls{CSI} feedback bottleneck for large $M$.
This limitation is addressed in the literature by decomposing the optimal precoder into the baseband and the RF precoders and by sending the codebook indices or quantized values of the precoders, which have much lower dimensions than the optimal precoder, over the feedback link \cite{el2014spatially,love2004value,alkhateeb2015limited}, and \cite{Ren2017}. 
Even though this approach reduces the excessive feedback overhead in massive \gls{MIMO} systems and makes it dependent on the number of transmit RF chains and data streams rather than the number of transmit antennas, a precoder design that can be adapted according to the actual constraint on the rate of the feedback link has not been proposed yet. Therefore, the goal of this paper is to design a precoder whose achievable rate can be traded for the affordable feedback overhead. To this end, we will exploit the special structure of the \gls{mmWave} channel, as explained in the next section.

%% file: sections/Feedback2.tex
\section{Feedback-Aware Precoder Design}\label{sec:feedback}

In this section, we design an adaptive precoder for \gls{mmWave} massive \gls{MIMO} systems. We first discuss the proposed design methodology and then analyze the corresponding feedback overhead. 

\subsection{Proposed Precoder Design}\label{subsec:precoder}
We start by examining the structure of the optimal unconstrained precoder. Let $\mat H=\mathbf{U}\boldsymbol{\Sigma}\mathbf{V}\hermis$ denote the singular value decomposition (SVD) of the channel matrix $\mathbf{H}$, where $\mathbf{U}$ and $\mathbf{V}$ are unitary matrices containing the left and right singular vectors, respectively, and $\boldsymbol{\Sigma}$ is a diagonal matrix containing the singular values. Thereby, the optimal linear precoder that maximizes the mutual information between $\mat s$ and $\mat y$ is $\Fopt= [\alpha_1\mathbf{v}_1,\dots,\alpha_s\mathbf{v}_S]$, where column vector $\mat{v}_s\in \set C^{M\times1}$ is the right singular vector corresponding to the $s$-th largest singular value of $\mat H$ and $\alpha_s$ is the power allocation factor for the $s$-th stream that has to satisfy $\sum_s \alpha_s^2=P$ and can be obtained with the water-filling algorithm~\cite{biglieri2007mimo}.
Additionally, given the clustered channel model in \eqref{Eq:Channel}, we can give a compact representation \mbox{$\mat H = \mat H_r \bar{\mat H} \mat H_t\hermis$},
where $\bar{\mat H}$ is a diagonal matrix containing the coefficients of all rays/path in all clusters,
and $\mat H_r = \big[\mathbf{h}_r(\theta_{1}),\dots,\mathbf{h}_r(\theta_{LJ})\big]\in \set C^{N\times LJ}$ and
\mbox{$\mat H_t=\big[\mathbf{h}_t(\phi_{1}),\dots,\mathbf{h}_t(\phi_{LJ})\big] \in \set C^{M\times LJ}$} are the receiver and transmitter array response matrices for corresponding \glspl{AoA} and \glspl{AoD}, respectively.
Given the previous observations, it can be shown that the optimal precoder can be written as $\Fopt=\sum_{k=1}^{LJ} \mat h_t(\phi_k)\mat{g}_k\hermis$, where $\phi_k$ is the $k$-th channel \gls{AoD} and $\mat{g}_k\in\set C^{S\times 1}$ is a corresponding appropriate combining vector \cite{el2014spatially}.

Based on the above, we propose to approximate the optimal precoder as follows
\begin{IEEEeqnarray}{lll} \label{Eq:precoder}
& \underset{\boldsymbol{\phi} \in \Cphi^{K\times 1}, \mat{G}\in\set C^{K\times S}}{\mathrm{minimize}} \frob{\Fopt-\CHt(\boldsymbol{\phi}) \mat{G}},\\
& \text{subject to: } \frob{\boldsymbol{\Psi}(\boldsymbol{\phi})\mat{G}}=1,	\nonumber
\end{IEEEeqnarray}
%
%
where $\mat{G}\in\set C^{K\times S}$ is the combining matrix and $\CHt(\boldsymbol{\phi})\in\set C^{M\times K}$ is the basis matrix whose columns are the transmitter array response evaluated at the discrete \glspl{AoD} in $\boldsymbol{\phi} = [\hat \phi_1,\dots,\hat \phi_\K] \transp \in \Cphi^{K\times 1}$, i.e., 
\begin{IEEEeqnarray}{lll} \label{Eq:CHt}
	\CHt(\boldsymbol{\phi})=\big[\mathbf{h}_t(\hat \phi_1),\dots,\mathbf{h}_t(\hat \phi_{K})\big].
\end{IEEEeqnarray}
Moreover, $\Cphi$ denotes the set of discrete angles used to approximate the \glspl{AoD} $\phi_k,\,\,\forall k$, and $\K$ is a design parameter and determines the number of \glspl{AoD} used for approximating the optimal precoder.  Since in general the allowed feedback overhead is small, we choose $\K\ll LJ$. Therefore, the problem in \eqref{Eq:precoder} is equivalent to the sparse recovery problem in compressive sensing. Thereby, \gls{OMP} is a widely-adopted algorithm, which can efficiently find the best $\K$ elements of basis matrix $\boldsymbol{\Psi}(\mathcal{C}_{\phi})\in\mathbb{C}^{M\times|\mathcal{C}_{\phi}|}$ to approximate $\mat{F}^{\mathrm{opt}}$~\cite{tropp2007signal}. Once $\boldsymbol{\phi}$ is derived, the optimization problem in \eqref{Eq:precoder} reduces to the well-known least square problem for $\mat G$ whose closed-form solution is known. Algorithm~1 summarizes the proposed precoder design based on \gls{OMP} and returns solutions $\phiopt$ and $\Gopt$.

\begin{algorithm}[t]
 	\caption{Feedback-Aware Precoding via \gls{OMP}}
 	\textbf{Require:} $\Fopt, \Cphi$
 	\begin{algorithmic}
 		\State 1: $\boldsymbol{\phi} = \text{Empty vector}$
 		\State 2: $\mat F_\text{res} = \Fopt$
 		\For{$i=1:\K$}
 		\State 4: $\mat Q = \CHt\hermis(\Cphi) \mat F_\text{res}$
 		\State 5: $\hat \phi = {\mathrm{argmax}}_{l=1:|\Cphi|} (\mat Q \mat Q\hermis)_{l,l} $
 		\State 6: $\boldsymbol{\phi} = [\boldsymbol{\phi} , \hat \phi]$
 		\State 7: $\mat G = \big(\CHt\hermis(\boldsymbol{\phi}) \CHt(\boldsymbol{\phi})\big)^{-1}  \CHt\hermis(\boldsymbol{\phi}) \Fopt  $
 		\State 8: $\mat F_\text{res} = { {\Fopt-\CHt(\boldsymbol{\phi})\mat{G}} \over \frob{\Fopt-\CHt(\boldsymbol{\phi})\mat{G}} }$
 		\EndFor
 		\State 9: $\phiopt=\boldsymbol{\phi}$ and $\Gopt =  {\mat{G} \over \frob{\CHt(\boldsymbol{\phi})\mat{G}}}$
 		\State \textbf{Return} $\phiopt, \Gopt $
	\end{algorithmic} 
\end{algorithm}

Note that the transmitter only needs to know $\phiopt$ and $\Gopt$ to construct the precoder $\hat{\mat{F}}=\CHt(\phiopt)\Gopt$. Next, the transmitter can fully implement this precoder based on Lemma~1 for the hybrid MIMO architecture introduced in Section~\ref{sec:sysmod}.
The structure of precoder $\hat{\mathbf{F}}$ implies that the required feedback scales with $\K(1+S)$, which is not only independent of the number of transmit antennas $M$, but can also be controlled with the proposed design parameter $\K$.
Therefore, the solution of optimization problem \eqref{Eq:precoder} ensures that the set of the best $\K$ vectors for spanning the channel's row space and their corresponding combining matrix are fed back to the transmitter.
Consequently, the optimal precoder can be better approximated by choosing a larger $\K$ at the cost of increased feedback overhead, while choosing a smaller $\K$ can lower the overhead when the feedback link bandwidth is limited.
We will study the effect of parameter $K$ on the precoding performance in Section~\ref{sec:sim}.


\subsection{Extension to Multi-Beam Basis Elements}\label{subsec:non-modulo-one}
The columns of the matrix $\CHt(\boldsymbol \phi)$ defined in \eqref{Eq:CHt} are the transmitter array response vectors at the angles in $\boldsymbol \phi$.
Recall that the angles in $\boldsymbol \phi$ are selected from the channel \gls{AoD} codebook $\Cphi$, where the transmitter sector is partitioned into sub-sectors by the codebook resolution and the middle angle of each sub-sector is naturally selected as a codebook element.
Therefore, the higher the codebook resolution is, the better the beams of $\CHt(\Cphi)$ can approximate all possible \gls{AoD} realizations.
However, if the codebook resolution is forced to be low due to the limited affordable feedback overhead, the beams of $\CHt(\Cphi)$ cannot cover all the \gls{AoD} realizations in the transmitter beam sector and thus the performance of $\hat{\mat{F}}$ can be severely degraded.
Although this is a fundamental limitation of any precoder design scheme, it can be remedied for the proposed precoder.

Note that the MIMO architecture presented in Section~\ref{sec:sysmod} can implement any precoder matrix and hence the decomposition in \eqref{Eq:precoder} does not impose any implementation constraint on the basis matrix $\CHt(\boldsymbol{\phi})$.
Therefore, the structure of $\CHt(\boldsymbol \phi)$ only needs to be agreed to by the transmitter and the receiver.
Motivated by the above, we propose to redefine $\CHt(\boldsymbol{\phi})$ in \eqref{Eq:precoder} and Algorithm~1 as
\begin{IEEEeqnarray}{lll} \label{Eq:non_modulo1_array}
	\CHt(\boldsymbol{\phi})={1 \over \sqrt{\Gamma}} \Big[\sum_{\gamma=1}^{\Gamma}&\mathbf{h}_t\left(\hat \phi_1-\frac{\Delta \phi}{2}+\frac{\gamma\Delta \phi}{\Gamma+1}\right),
	\dots, \nonumber \\ 
	&\sum_{\gamma=1}^{\Gamma}\mathbf{h}_t\left(\hat \phi_K-\frac{\Delta \phi}{2}+\frac{\gamma\Delta \phi}{\Gamma+1}\right)  \Big],
\end{IEEEeqnarray}
where $\Delta\phi$ is the resolution of the \glspl{AoD} in $\Cphi$ and $\Gamma$ is a design parameter. This proposition implies that the beam candidate of each sub-sector is a superposition of $\Gamma$ beams of the angles in that sub-sector. For $\Gamma=1$, the beam candidate is the array response vector at the middle angle of the corresponding sub-sector and thus \eqref{Eq:non_modulo1_array} reduces to \eqref{Eq:CHt}.
Note that the proposition in \eqref{Eq:non_modulo1_array} does not impose any extra feedback overhead as $\Gamma$ is agreed to by the transmitter and the receiver. 

\begin{figure}[t]
	\centering
	\input{./fig/multi_beam.tex}
	\caption{$g_k(\phi)$ vs. $\phi$ for $\hat \phi_k=0$ and $\Gamma\in\{1,2,4\}$. The angles between the black vertical lines constitute the sub-sector that should be covered by the basis element $\mat{\Psi}(\hat \phi_k)$.}
	\label{fig:beam}
\end{figure}
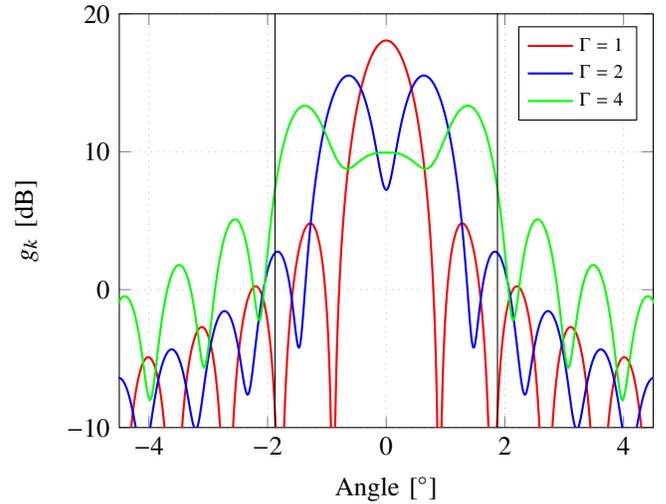

Examples for beam patterns obtained with the proposed approach are shown in Fig.~\ref{fig:beam}, where $M=128$, $|\Cphi|=16$, and a sector covering $\phi\in[-\pi/6,\pi/6]$ are assumed. In particular, in Fig.~\ref{fig:beam}, we show $g_k(\phi)$ versus $\phi$ where $g_k(\phi)$ is defined as 
\begin{IEEEeqnarray}{lll}
	g_k(\phi) = \frac{|\mathbf{h}_t^*(\phi)\CHt(\hat \phi_k)|^2}{\int_{\phi} |\mathbf{h}_t^*(\phi)\CHt(\hat \phi_k)|^2\mathrm{d}\phi},
\end{IEEEeqnarray}
characterizing the normalized power radiated in direction $\phi$ when $\CHt(\hat \phi_k)$ is adopted as the RF beamformer.
As can be observed from Fig.~\ref{fig:beam}, for $\Gamma= 1$, the power distribution across the sub-sector significantly varies with the angle and there are even several nulls within the sub-sector. This leads to a severe performance degradation when the actual \glspl{AoD} are at or close to these nulls. On the other hand, the power distribution across the sub-sector becomes more uniform when $\Gamma$ is increased. Nevertheless, by increasing $\Gamma$, the maximum achievable beamforming gain decreases and the interfering side lobes increase, too. 
Therefore, $\Gamma$ is also a design parameter that should be properly chosen for a given \gls{AoD} codebook resolution $|\Cphi|$. We will study the effect of this parameter on precoder performance in Section~\ref{sec:sim}.

\subsection{Feedback Overhead Quantification}

In the following, we quantify the feedback overhead of several benchmark precoder designs reported in the literature and the proposed precoder. Subsequently, for some special cases, we provide insightful results for the required feedback overhead of the proposed precoder.

\subsubsection{Comparison} In general, the precoder can be computed at the transmitter or the receiver. In the latter case, the precoder obtained at the receiver has to be fed back to the transmitter whereas in the former case, the CSI estimated at the receiver has to be fed back to the transmitter.
In the following, we review the corresponding feedback overhead requirements.

\textbf{Direct feedback:} We consider the case where each element of the channel matrix is quantized and fed back to the transmitter, i.e., individual quantized feedback, and the transmitter computes the precoder based on the acquired CSI knowledge. Let $\mathcal{C}_{c}$ denote the set of complex numbers used for quantization. Thereby, the overall feedback overhead for this scheme is $MN\log_2|\mathcal{C}_{c}|$ bits. Alternatively, the receiver may compute the precoder and quantize each element of the precoder and feed them back to the transmitter, which leads to an overall feedback requirement of $MS\log_2|\mathcal{C}_{c}|$ bits. Note that the latter scheme requires lower feedback overhead as $S\leq N$ holds.

\textbf{Sparse precoder designs:} Exploiting the sparsity of the mmWave channel, precoders of different complexity are designed in \cite{el2014spatially} and \cite{rusu2016low} for the fully-connected hybrid MIMO architecture, which employs $Q$ RF chains where $Q\geq S$. The overall feedback requirement of these precoder designs is $Q \mathrm{log_2}|\Cphi| + Q S \,\mathrm{log_2}|\mathcal{C}_{c}|$ bits where the terms $Q \mathrm{log_2}|\Cphi|$ and $Q S \,\mathrm{log_2}|\mathcal{C}_{c}|$ correspond to the analog and digital precoders, respectively. Note that unlike the feedback overhead of the direct schemes, the overhead of the precoder designs in \cite{el2014spatially} and \cite{rusu2016low} does not scale with the number of antennas $M$.

\textbf{Multi-level channel estimation:} mmWave channel acquisition is considered in \cite{hur2013millimeter} and \cite{alkhateeb2014channel} and multi-level RF codebooks are designed, which are able to improve the CSI acquisition quality by increasing the adopted number of codebook levels. Thereby, for FDD systems, quantized versions of the $K$ strongest AoDs, $\hat{\phi}_{1},\dots,\hat{\phi}_{K}$, and AoAs, denoted by $\hat{\theta}_{1},\dots,\hat{\theta}_{K}$,
and their corresponding channel path coefficients, denoted by $\hat{\bar{h}}_{1},\dots,\hat{\bar{h}}_{K}$, are fed back to the transmitter. The transmitter is then able to reconstruct the channel matrix as $\hat{\mathbf{H}}=\hat{\mat{H}}_r \hat{\bar{\mat H}} \hat{\mat{H}}_t\hermis$,  
where $\hat{\mat{H}}_r = \big[\mathbf{h}_r(\hat{\theta}_{1}),\dots,\mathbf{h}_r(\hat{\theta}_K)\big]$ and
$\hat{\mat{H}}_t=\big[\mathbf{h}_t(\hat{\phi}_{1}),\dots,\mathbf{h}_t(\hat{\phi}_{K})\big]$, and 
$\hat{\bar{\mat H}}$ is a diagonal matrix with non-zero elements $\hat{\bar{h}}_{1},\dots,\hat{\bar{h}}_{K}$. Having the estimated channel  $\hat{\mathbf{H}}$, the transmitter computes the desired precoder. The feedback overhead of this scheme is $2K \mathrm{log_2}|\Cphi|+ K \,\mathrm{log_2}|\mathcal{C}_c|$ where we assume the same codebook $\Cphi$ for both the AoAs and AoDs for simplicity. Note that this scheme has the advantage that the accuracy of CSI acquisition and correspondingly the feedback overhead requirement can be adjusted by parameter $K$.

\textbf{Proposed precoder design:} Similar to \cite{el2014spatially} and \cite{rusu2016low}, we exploit the sparsity of the mmWave channel; however, we developed our precoder based on the hybrid MIMO architecture in \cite{zhang2014achieving}. The overall feedback requirement of our precoder design is $K \mathrm{log_2}|\Cphi| + K S \,\mathrm{log_2}|\mathcal{C}_{c}|$ bits. Note that unlike the fixed feedback overhead of the precoder designs in \cite{el2014spatially} and \cite{rusu2016low}, the feedback overhead of the proposed design can be adjusted by choosing the parameter $K$. This enables the proposed design to adapt itself to a feedback overhead that can be afforded by the system. Unlike \cite{hur2013millimeter} and \cite{alkhateeb2014channel}, which compute the precoder at the transmitter, we compute the precoder at the receiver, which is shown in Section~IV to be more efficient.

Table~\ref{tab:comparison} summarizes the overhead for different precoder designs for point-to-point MIMO systems.

\begin{table}[t]
	\vspace{3mm}
	\caption{Feedback overhead comparison}\label{tab:comparison}
	\centering
	\resizebox{\columnwidth}{!}
	{	
		\begin{tabular}{|c|c|c|}
			\hline
			\Xhline{2\arrayrulewidth}
			Precoding scheme & \multicolumn{2}{c|}{Total feedback overhead [bits]} \\
			\hline
			\hline
			Feeding back elements of $\mathbf{H}$ & \multicolumn{2}{c|}{$MN\log_2|\mathcal{C}_{c}|$} \\ \hline 
			Feeding back elements of $\mathbf{F}$ & \multicolumn{2}{c|}{$MS\log_2|\mathcal{C}_{c}|$} \\ \hline
			\multicolumn{1}{c|}{ } & Angles & \multicolumn{1}{c|}{Amplitudes} \\\hline
			Sparse precoder design \cite{el2014spatially,rusu2016low} & $Q \mathrm{log_2}|\Cphi|$ & $Q S \,\mathrm{log_2}|\mathcal{C}_{c}|$\\\hline
			Multi-level channel estimation \cite{hur2013millimeter,alkhateeb2014channel} &  $2K \mathrm{log_2}|\Cphi|$ & $K\mathrm{log_2}|\mathcal{C}_c|$\\ \hline
			Proposed precoder design  & $K \mathrm{log_2}|\Cphi|$ & $K S\mathrm{log_2}|\mathcal{C}_c|$\\
			\hline
		\end{tabular}
	}
\end{table}

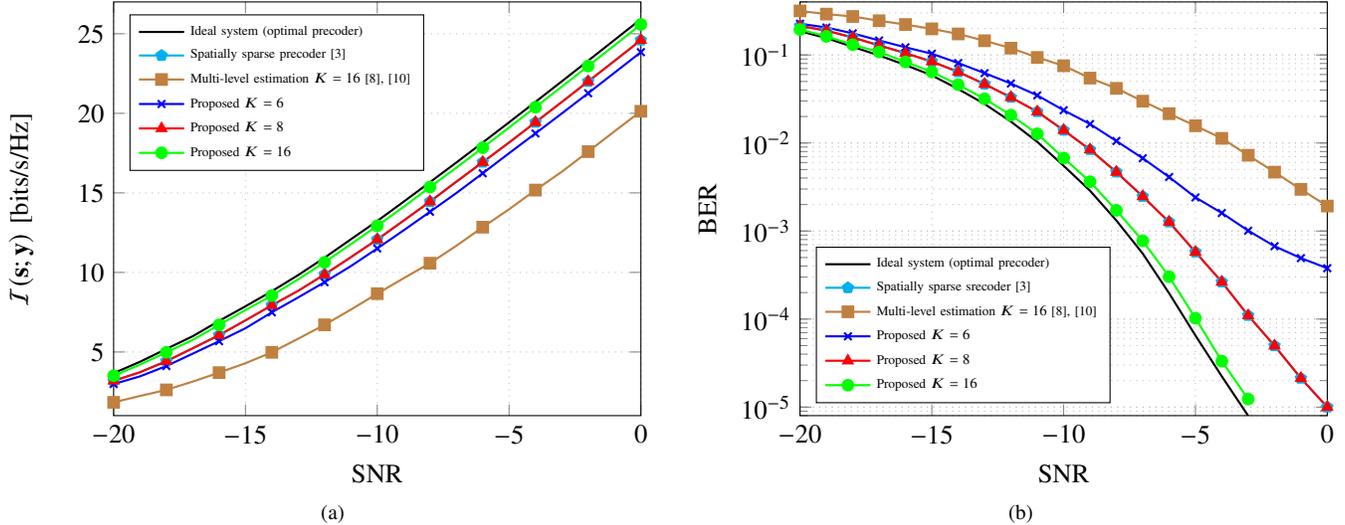
\begin{figure*}[!t]
	\centering
	\subfloat[\label{subfig-1}]{\input{./fig/rate2.tex}} \hspace{2mm}
	\subfloat[\label{subfig-2}]{\input{./fig/ber1.tex}} 
	\caption{a) Achievable rate $\mathcal{I}(\mat s;\mat y)$; and b) BER of the proposed precoding scheme with $\K=6, 8,$ and $16$ in comparison with the ideal system with the optimal precoder $\Fopt$, the spatially sparse precoding scheme presented in \cite{el2014spatially}, and the precoder based on the multi-level channel estimation presented in \cite{hur2013millimeter} and \cite{alkhateeb2014channel}.} 
	\label{fig:ber}
	\vspace{-3mm}
\end{figure*}

\subsubsection{Scaling Order of the Feedback Overhead for the Proposed Precoder Design}
We first present the scaling order of the required feedback overhead of the proposed precoder needed to approach the performance of an \emph{ideal} system, where no feedback constraint is applied, on the implementation of the optimal precoder.
We further provide the scaling overhead of the required feedback for asymptotic case when $M\to\infty$.

\begin{lem}
	To approach the performance of the ideal system, the feedback overhead requirement for the proposed precoding scheme is on the order of $LJ(\mathrm{log_2}|\Cphi|+S\mathrm{log_2}|\mathcal{C}_c|)$ bits assuming that $|\Cphi|$ and $|\mathcal{C}_c|$ are sufficiently large.	
\end{lem}
\begin{IEEEproof}
	For the non-asymptotic regime of $M$, the columns of the optimal unitary matrix $\Fopt$ may depend on the array response of all existing \glspl{AoD}. Therefore, we choose $\K=LJ$. Thereby, $\hat{\mathbf{F}}$ approaches $\Fopt$ assuming that $|\Cphi|$ and $|\mathcal{C}_c|$ are sufficiently large. This leads to an overhead requirement of $LJ(\mathrm{log_2}|\Cphi|+S\mathrm{log_2}|\mathcal{C}_c|)$ bits and concludes the proof.
\end{IEEEproof}

The above lemma states that as long as the \gls{mmWave} channel contains limited scattering, i.e., $LJ\ll M$, the proposed scheme can effectively approach the performance of the ideal system with a limited feedback requirement that linearly scales with $LJ$.  


\begin{lem}
	Asymptotically as $M\to\infty$, the feedback overhead requirement for the proposed precoding scheme to approach the performance of the ideal system is on the order of $S(\mathrm{log_2}|\Cphi|+\mathrm{log_2}|\mathcal{C}_c|)$ bits assuming that $|\Cphi|$ and $|\mathcal{C}_c|$ are sufficiently large.
\end{lem}
\begin{IEEEproof}
	As $M\to\infty$, the columns of matrix $\mat{H}_t$ become orthogonal~\cite{el2012capacity}. Thereby, the columns of the optimal matrix $\Fopt$ become identical to a scaled version of the first $S$ columns of $\mat{H}_t$, where the scaling factors are the power allocation variables. Therefore, assuming $\K \geq S$, matrix $\Fopt$ can be fully reconstructed from (\ref{Eq:precoder}) by choosing $K=S$ and $\Gopt= \text{diag}[\alpha_1,\dots,\alpha_S]$.
	Thereby, assuming that $|\Cphi|$ and $|\mathcal{C}_c|$ are sufficiently large, the feedback overhead is on the order of $S(\mathrm{log_2}|\Cphi|+\mathrm{log_2}|\mathcal{C}_c|)$ bits. This completes the proof.
\end{IEEEproof}

The above lemma states that asymptotically as $M\to\infty$, simple beam steering along the $S$ dominant \gls{AoA} is optimal.

%% file: fig/multi_beam.tex
\begin{tikzpicture}
\pgfplotsset{grid style={dotted}}
        \begin{axis}[
        width = 0.98\columnwidth,
        height = 0.80\columnwidth,
        xlabel = {Angle [\degree]},
        ylabel = {$g_k$ [dB]},
        xmin = -4.5, xmax = 4.5,
        ymin = -10, ymax = 20,
        grid = both,
        legend style={legend pos=north east,font=\scriptsize},
        legend cell align=left,
        legend entries={$\Gamma=1$, $\Gamma=2$, $\Gamma=4$},
        ]		
	\addplot[red, thick, solid] table[x index=0,y index=1]{./fig/data/gk_gamma1.dat};
	\addplot[blue, thick, solid] table[x index=0,y index=1]{./fig/data/gk_gamma2.dat};
	\addplot[green, thick, solid] table[x index=0,y index=1,each nth point={5}]{./fig/data/gk_gamma4.dat};
	\draw (axis cs:-1.875,-15) -- (axis cs:-1.875,25);
	\draw (axis cs:1.875,-15) -- (axis cs:1.875,25);
	\end{axis}
\end{tikzpicture}%

%% file: fig/rate2.tex
\begin{tikzpicture}	
\pgfplotsset{grid style={dotted}}
        \begin{axis}[
        width = 0.97\columnwidth,
        height = 0.80\columnwidth,
        xlabel = {SNR},
        ylabel = {$\mathcal{I}(\mat s;\mat y)$ [bits/s/Hz]},
        xmin = -20, xmax = 0, 
        ymin = 1, ymax = 27, 
        grid = both,
        legend style={legend pos=north west,font=\tiny},
        legend cell align=left,
        legend entries={Optimal Precoder, Spatially Sparse Precoder, Proposed $K=6$, Proposed $K=8$, Proposed $K=16$},
        name=plot2,
        ]
       	\addlegendimage{black, thick, solid};	
       	\addlegendentry{Ideal system (optimal precoder)};
       	
       	\addlegendimage{cyan, thick,solid, mark=pentagon*};	
       	\addlegendentry{Spatially sparse precoder \cite{el2014spatially}};
       	
       	\addlegendimage{brown, thick,solid, mark=square*};
       	\addlegendentry{Multi-level estimation $K=16$ \cite{hur2013millimeter,alkhateeb2014channel}};
       	
       	\addlegendimage{blue, thick,solid, mark=x};
       	\addlegendentry{Proposed $K=6$};
       	
       	\addlegendimage{red, thick,solid, mark=triangle*};
       	\addlegendentry{Proposed $K=8$};
       	
       	\addlegendimage{green,thick,solid, mark=*};
       	\addlegendentry{Proposed $K=16$};
       	
        			
	\addplot[black, thick, solid] table[x index=0,y index=1]{./fig/data/rate_optimal_12cl.dat};			
	\addplot[cyan, thick, solid, mark=pentagon*,  mark repeat=2] table[x index=0,y index=1]{./fig/data/rate_OMP_12cl.dat};
	\addplot[brown, thick, solid, mark=square*,  mark repeat=2] table[x index=0,y index=1]{./fig/data/rate_optimal_chEstimate16_12cl.dat};	
	\addplot[blue, thick, solid, mark=x,  mark repeat=2] table[x index=0,y index=1]{./fig/data/rate_FoptDec_12cl_N6.dat};        
	\addplot[red, thick, solid, mark=triangle*,  mark repeat=2] table[x index=0,y index=1]{./fig/data/rate_FoptDec_12cl_N8.dat}; 
	\addplot[green, thick, solid, mark=*,  mark repeat=2] table[x index=0,y index=1]{./fig/data/rate_FoptDec_12cl_N16.dat};      
	
        \end{axis}
\end{tikzpicture}

%% file: fig/ber1.tex
\begin{tikzpicture}

\pgfplotsset{grid style={dotted}}

	\begin{semilogyaxis}[
	width = 0.97\columnwidth,
	height = 0.80\columnwidth,
	xlabel = {SNR},
	ylabel = {BER},
	xmin = -20, xmax = 0, 
	ymin = 8e-6, ymax = 4e-1,
	grid = both,
	legend style={legend pos=south west,font=\tiny},
	legend cell align=left,
	name=plot1,
	at=(plot2.right of south east),xshift=15mm
	]
	
	\addlegendimage{black, thick, solid}; 
	\addlegendentry{Ideal system (optimal precoder)};
	
	\addlegendimage{cyan, thick,solid, mark=pentagon*};	
	\addlegendentry{Spatially sparse srecoder \cite{el2014spatially}};
	
	\addlegendimage{brown, thick,solid, mark=square*};
	\addlegendentry{Multi-level estimation $K=16$ \cite{hur2013millimeter,alkhateeb2014channel}};
	
	\addlegendimage{blue, thick,solid, mark=x};
	\addlegendentry{Proposed $K=6$};
	
	\addlegendimage{red, thick,solid, mark=triangle*};
	\addlegendentry{Proposed $K=8$};
	
	\addlegendimage{green, thick,solid, thick,  mark=*};
	\addlegendentry{Proposed $K=16$};
	
	
	\addplot[black, thick, solid] table[x index=0,y index=1]{./fig/data/optimal_12cl.dat};			
	\addplot[cyan, thick, solid, mark=pentagon*] table[x index=0,y index=1]{./fig/data/OMP_12cl.dat};	
	\addplot[brown, thick, solid, mark=square*] table[x index=0,y index=1]{./fig/data/optimal_chEstimate16_12cl.dat};
	\addplot[blue, thick, solid, mark=x] table[x index=0,y index=1]{./fig/data/FoptDec_12cl_N6.dat};        
	\addplot[red, thick, solid, mark=triangle*] table[x index=0,y index=1]{./fig/data/FoptDec_12cl_N8.dat}; 
	\addplot[green, thick, solid, mark=*] table[x index=0,y index=1]{./fig/data/FoptDec_12cl_N16.dat};      

	
	\end{semilogyaxis}
\end{tikzpicture}

%% file: sections/SimulationResults.tex
\section{Comparison and Simulation Results}\label{sec:sim}


In this section, we demonstrate the performance of the proposed precoder to illustrate the impact of the design parameters $\K$, $\Gamma$, and $|\Cphi|$.

\subsection{Simulation Setup and Benchmarks}
We assume $M=128$ transmitter antennas, $N=16$ receive antennas, and $S=4$ independent data streams. We consider the channel model described in \eqref{Eq:Channel} with $L=12$ clusters and $J=20$ rays per cluster. The channel coefficients $h_{k}$ are independent and identically distributed normal random variables $\mathcal{CN}(0,1)$ and the \glspl{AoD} (\glspl{AoA}) within each cluster $\phi_{k}$ ($\theta_{k}$) are distributed according to a Laplacian distribution with a \mbox{uniformly-random} mean within a $90\degree$-wide transmitter ($360\degree$-wide receiver) sector. We also assume a uniform linear array with $d = \lambda / 2$ at both the transmitter and receiver.

We consider three benchmark schemes. For all benchmark schemes, we assume ideal channel estimation at the receiver and focus on the required CSI feedback overhead for the precoder design at the transmitter.

\textbf{Benchmark~1:}
As the first benchmark scheme, we adopt the precoder in \cite{el2014spatially} which was designed to approximate  the optimal unconstrained unitary precoder, i.e., $\Fopt$ with $\alpha_1=\dots=\alpha_S$. Therefore, for a fair comparison, we also use the optimal unitary precoder  for the design of our proposed precoder in (\ref{Eq:precoder}).
\footnote{We note that the proposed precoder design is general and can also be used when the $\alpha_s$ are not identical.}
Further, we consider $Q=8$ transmit RF chains for this benchmark scheme. Therefore, from a hardware perspective, the considered architecture in Fig.~\ref{fig:SysMod} is simpler than that assumed for this benchmark scheme since it has half the number of RF chains and the same number of PSs, i.e., $8\times 128$.

\textbf{Benchmark~2:}
As the second benchmark scheme, we consider the case when the multi-level estimated channel is fed back to the transmitter \cite{hur2013millimeter} and \cite{alkhateeb2014channel}.
Then, the optimal unitary precoder is computed based on the SVD of the fed back CSI, which can be realized using the architecture in Fig.~\ref{fig:SysMod}.

\textbf{Benchmark~3 (Upper bound):}
We assume the ideal system for this benchmark, and therefore, we consider the optimal unitary precoder computed based on the exact CSI at the transmitter as a performance upper bound.

Note that we show the results for i) coded transmission in terms of the achievable rate $\mathcal{I}(\mat s;\mat y)=\text{log}_2\big(\big|\mat{I}_N + {P \over \sigma_n^2} \mat{H} \mat{F} \mat{F}\hermis \mat{H}\hermis\big| \big)$, where $\mathcal{I}(\mat s;\mat y)$ denotes the mutual information between $\mat s$ and $\mat y$ and ii) uncoded transmission with \gls{QPSK} modulation and linear \gls{MMSE} detection in terms of the \gls{BER}. Due to space constraints, we assume ideal feedback for the amplitudes and focus on the feedback overhead of the angles and the impact of parameters $\K$, $\Gamma$, and $|\Cphi|$.

\subsection{Simulation Results}
Fig.~\ref{fig:ber} shows the performance of the proposed precoding scheme for three feedback parameters $\K=6, 8,$ and $16$. 
We plot $\mathcal{I}(\mat s;\mat y)$ in Fig.~\ref{subfig-1} and  \gls{BER} in Fig.~\ref{subfig-2}  vs. the \gls{SNR} defined as $P \over \sigma_n^2$.
A sufficiently large resolution for the \gls{AoD} codebook, i.e., $|\Cphi|=2^8$, and $\Gamma=1$ are assumed, to study the impact of $K$.
The results in Fig.~\ref{fig:ber} indicate that by changing parameter $\K$, the precoder performance can be traded with feedback overhead. Specifically, by choosing \mbox{$\K=8$}, the proposed precoder achieves the same performance and requires the same amount of feedback overhead as the precoder presented in~\cite{el2014spatially}.\footnote{Note that by choosing $\K$ equal to number of RF chains in \cite{el2014spatially}, $\Gopt$ and $\CHt(\phiopt)$ in \eqref{Eq:precoder} will be equal to $\fbb$ and $\frf$ defined in \cite[Eq. (16)]{el2014spatially}, respectively.}
However, by increasing $\K$ to $16$ and doubling the feedback overhead, the performance of the proposed precoder is improved by almost $3$~dB in terms of BER and $2$~dB in terms of achievable rate and closely approaches the performance of the optimal precoder.
Furthermore, for $\K=6$, the feedback overhead is decreased at the expense of a lower precoder performance.
It is worth noting that, although  \cite{alkhateeb2014channel} provides the flexibility of estimating the channel with different numbers of paths and adapting the feedback rate, for similar feedback overhead, the corresponding performance is much lower than that of the proposed precoder design.

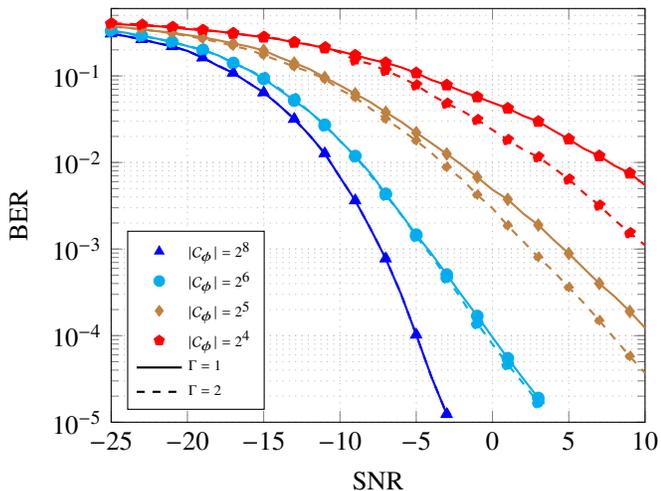
\begin{figure}[t]
	\centering
	\input{./fig/ber_resolution.tex}
	\caption{BER of the proposed precoding scheme versus SNR for different values of $|\Cphi|$ and $\Gamma$.}
	\label{fig:resolution}
	\vspace{-3mm}
\end{figure}

In Fig.~\ref{fig:resolution}, we evaluate the impact of 
parameter $\Gamma$ for $K=16$ on the \gls{BER}. The figure shows the precoder performance for $|\Cphi|=2^4, 2^5, 2^6,$ and $2^8$ with $\Gamma=1$ and $\Gamma=2$. As can be observed, increasing $\Gamma$ can improve the precoder performance in the low resolution regime, i.e., for $2^4$ and $2^5$, and partially compensate for the corresponding performance loss due the insufficient codebook resolution, while this improvement becomes small for higher resolutions, i.e., for $2^6$ and $2^8$.

%% file: fig/ber_resolution.tex
\begin{tikzpicture}
\pgfplotsset{grid style={dotted}}
        \begin{semilogyaxis}[
        width = 0.98\columnwidth,
        height = 0.80\columnwidth,
        xlabel = {SNR},
        ylabel = {BER},
        xmin = -25, xmax = 10,
        ymin = 1e-5, ymax = 6e-1,
        grid = both,
        legend style={legend pos=south west,font=\tiny},
        legend cell align=left,
        ]
        
        \addlegendimage{blue, only marks, mark=triangle*};
	\addlegendentry{$|\Cphi|=2^8$};
	\addlegendimage{cyan, only marks, mark=*};
	\addlegendentry{$|\Cphi|=2^6$};
	\addlegendimage{brown, only marks, mark=diamond*};
	\addlegendentry{$|\Cphi|=2^5$};
	\addlegendimage{red, only marks, mark=pentagon*};
	\addlegendentry{$|\Cphi|=2^4$};
	
	\addlegendimage{black,thick,solid};
        \addlegendentry{$\Gamma=1$}
        \addlegendimage{black,thick,dashed};
        \addlegendentry{$\Gamma=2$}
	     	
	\addplot[blue, thick, solid, mark=triangle*, mark repeat=2] table[x index=0,y index=1]{./fig/data/Res256_simple.dat}; 	
	\addplot[cyan, thick, solid, mark=*, mark repeat=2] table[x index=0,y index=1]{./fig/data/Res64_simple.dat}; 		
	\addplot[brown, thick, solid, mark=diamond*, mark repeat=2] table[x index=0,y index=1]{./fig/data/Res32_simple.dat}; 	
	\addplot[red, thick, solid, mark=pentagon*, mark repeat=2] table[x index=0,y index=1]{./fig/data/Res16_simple.dat}; 	
	     
	\addplot[blue, thick, dashed, mark=triangle*, mark repeat=2] table[x index=0,y index=1]{./fig/data/Res256_subSec2.dat};
	\addplot[cyan, thick, dashed, mark=*, mark repeat=2] table[x index=0,y index=1]{./fig/data/Res64_subSec2.dat};
	\addplot[brown, thick, dashed, mark=diamond*, mark repeat=2] table[x index=0,y index=1]{./fig/data/Res32_subSec2.dat};
	\addplot[red, thick, dashed, mark=pentagon*, mark repeat=2] table[x index=0,y index=1]{./fig/data/Res16_subSec2.dat};
	
	\end{semilogyaxis}
\end{tikzpicture}%

%% file: sections/Conclution.tex
\section{Conclusion}\label{sec:conc}
In this paper, we considered a narrow-band point-to-point massive \gls{MIMO} system with a hybrid architecture that employs two PSs for each coefficient of the analog precoder and is able to realize any arbitrary precoder. Exploiting the degrees of freedom offered by this architecture and the mmWave channel structure, we proposed a precoding scheme that approximates the optimal precoder by selecting a desired number of elements from the transmitter array response basis matrix and a corresponding combining matrix.
The proposed precoder design provided a flexible tradeoff between precoding performance and feedback overhead. 
Simulation results showed that the proposed precoding scheme can provide a near-optimal solution with a higher feedback overhead, while an acceptable approximation of the optimal precoder is maintained when a lower feedback overhead is demanded. Moreover, for a given affordable feedback overhead, the proposed precoder outperforms benchmark schemes from the literature.


%% file: main.bbl
\begin{thebibliography}{10}
\providecommand{\url}[1]{#1}
\csname url@samestyle\endcsname
\providecommand{\newblock}{\relax}
\providecommand{\bibinfo}[2]{#2}
\providecommand{\BIBentrySTDinterwordspacing}{\spaceskip=0pt\relax}
\providecommand{\BIBentryALTinterwordstretchfactor}{4}
\providecommand{\BIBentryALTinterwordspacing}{\spaceskip=\fontdimen2\font plus
\BIBentryALTinterwordstretchfactor\fontdimen3\font minus
  \fontdimen4\font\relax}
\providecommand{\BIBforeignlanguage}[2]{{%
\expandafter\ifx\csname l@#1\endcsname\relax
\typeout{** WARNING: IEEEtran.bst: No hyphenation pattern has been}%
\typeout{** loaded for the language `#1'. Using the pattern for}%
\typeout{** the default language instead.}%
\else
\language=\csname l@#1\endcsname
\fi
#2}}
\providecommand{\BIBdecl}{\relax}
\BIBdecl

\bibitem{han2015large}
S.~Han, I.~Chih-Lin, Z.~Xu, and C.~Rowell, ``Large-scale antenna systems with
  hybrid analog and digital beamforming for millimeter wave {5G},''
  \emph{{IEEE} Commun. Mag.}, vol.~53, no.~1, pp. 186--194, 2015.

\bibitem{zhang2005variable}
X.~Zhang, A.~F. Molisch, and S.-Y. Kung, ``Variable-phase-shift-based
  {RF}-baseband codesign for {MIMO} antenna selection,'' \emph{{IEEE} Trans.
  Signal Processing}, vol.~53, no.~11, pp. 4091--4103, 2005.

\bibitem{el2014spatially}
O.~El~Ayach, S.~Rajagopal, S.~Abu-Surra, Z.~Pi, and R.~W. Heath, ``Spatially
  sparse precoding in millimeter wave {MIMO} systems,'' \emph{{IEEE} Trans.
  Wireless Commun.}, vol.~13, no.~3, pp. 1499--1513, 2014.

\bibitem{yu2016alternating}
X.~Yu, J.-C. Shen, J.~Zhang, and K.~B. Letaief, ``Alternating minimization
  algorithms for hybrid precoding in millimeter wave {MIMO} systems,''
  \emph{{IEEE} J. Select. Topics Signal Processing.}, vol.~10, no.~3, pp.
  485--500, 2016.

\bibitem{rusu2016low}
C.~Rusu, R.~Mendez-Rial, N.~Gonz{\'a}lez-Prelcic, and R.~W. Heath, ``Low
  complexity hybrid precoding strategies for millimeter wave communication
  systems,'' \emph{{IEEE} Trans. Wireless Commun.}, vol.~15, no.~12, pp.
  8380--8393, 2016.

\bibitem{love2004value}
D.~J. Love, R.~W. Heath, W.~Santipach, and M.~L. Honig, ``What is the value of
  limited feedback for {MIMO} channels?'' \emph{{IEEE} Commun. Mag.}, vol.~42,
  no.~10, pp. 54--59, 2004.

\bibitem{wang2017hybrid}
H.~Wang, W.~Wang, V.~K. Lau, and Z.~Zhang, ``Hybrid limited feedback in {5G}
  cellular systems with massive {MIMO},'' \emph{IEEE Syst. Journal}, vol.~11,
  no.~1, pp. 50--61, 2017.

\bibitem{hur2013millimeter}
S.~Hur, T.~Kim, D.~J. Love, J.~V. Krogmeier, T.~A. Thomas, A.~Ghosh
  \emph{et~al.}, ``Millimeter wave beamforming for wireless backhaul and access
  in small cell networks.'' \emph{{IEEE} Trans. Commun.}, vol.~61, no.~10, pp.
  4391--4403, 2013.

\bibitem{noh2017multi}
S.~Noh, M.~D. Zoltowski, and D.~J. Love, ``Multi-resolution codebook and
  adaptive beamforming sequence design for millimeter wave beam alignment,''
  \emph{{IEEE} Trans. Wireless Commun.}, vol.~16, no.~9, pp. 5689--5701, 2017.

\bibitem{alkhateeb2014channel}
A.~Alkhateeb, O.~El~Ayach, G.~Leus, and R.~W. Heath, ``Channel estimation and
  hybrid precoding for millimeter wave cellular systems,'' \emph{{IEEE} J.
  Select. Topics Signal Processing.}, vol.~8, no.~5, pp. 831--846, 2014.

\bibitem{alkhateeb2015limited}
A.~Alkhateeb, G.~Leus, and R.~W. Heath, ``Limited feedback hybrid precoding for
  multi-user millimeter wave systems,'' \emph{{IEEE} Trans. Wireless Commun.},
  vol.~14, no.~11, pp. 6481--6494, 2015.

\bibitem{Ren2017}
Y.~Ren, Y.~Wang, C.~Qi, and Y.~Liu, ``Multiple-beam selection with limited
  feedback for hybrid beamforming in massive {MIMO} systems,'' \emph{IEEE
  Access}, vol.~5, pp. 13\,327--13\,335, 2017.

\bibitem{zhang2014achieving}
E.~Zhang and C.~Huang, ``On achieving optimal rate of digital precoder by
  {RF}-baseband codesign for {MIMO} systems,'' in \emph{Proc. Veh. Technol.
  Conf. (VTC Fall)}, 2014, pp. 1--5.

\bibitem{balanis1992antenna}
C.~A. Balanis, ``Antenna theory: A review,'' \emph{Proceedings of the IEEE},
  vol.~80, no.~1, pp. 7--23, 1992.

\bibitem{rappaport2014millimeter}
T.~S. Rappaport, R.~W. Heath~Jr, R.~C. Daniels, and J.~N. Murdock,
  \emph{Millimeter wave wireless communications}.\hskip 1em plus 0.5em minus
  0.4em\relax Pearson Education, 2014.

\bibitem{bogale2016number}
T.~E. Bogale, L.~B. Le, A.~Haghighat, and L.~Vandendorpe, ``On the number of
  {RF} chains and phase shifters, and scheduling design with hybrid
  analog--digital beamforming,'' \emph{{IEEE} Trans. Wireless Commun.},
  vol.~15, no.~5, pp. 3311--3326, 2016.

\bibitem{biglieri2007mimo}
E.~Biglieri, R.~Calderbank, A.~Constantinides, A.~Goldsmith, A.~Paulraj, and
  H.~V. Poor, \emph{{MIMO Wireless Communications}}.\hskip 1em plus 0.5em minus
  0.4em\relax Cambridge university press, 2007.

\bibitem{tropp2007signal}
J.~A. Tropp and A.~C. Gilbert, ``Signal recovery from random measurements via
  orthogonal matching pursuit,'' \emph{{IEEE} Trans. Inform. Theory}, vol.~53,
  no.~12, pp. 4655--4666, 2007.

\bibitem{el2012capacity}
O.~El~Ayach, R.~W. Heath, S.~Abu-Surra, S.~Rajagopal, and Z.~Pi, ``The capacity
  optimality of beam steering in large millimeter wave {MIMO} systems,'' in
  \emph{Proc. IEEE Workshop on Signal Processing Advances in Wireless Commun.
  (SPAWC)}, 2012, pp. 100--104.

\end{thebibliography}
